31 July, 2025

# Forecasting Binary Economic Events in Modern Mercantilism: Traditional methodologies coupled with PCA and K-means Quantitative Analysis of Qualitative Sentimental Data

Sebastian Kot

sebastian.kot@city.ac.uk

**Abstract**

This paper examines Modern Mercantilism, characterized by rising economic nationalism, strategic technological decoupling, and geopolitical fragmentation, as a disruptive shift from the post-1945 globalization paradigm. It applies Principal Component Analysis (PCA) to 768-dimensional SBERT-generated semantic embeddings of curated news articles to extract orthogonal latent factors that discriminate binary event outcomes linked to protectionism, technological sovereignty, and bloc realignments. Analysis of principal component loadings identifies key semantic features driving classification performance, enhancing interpretability and predictive accuracy. This methodology provides a scalable, data-driven framework for quantitatively tracking emergent mercantilist dynamics through high-dimensional text analytics

**Introduction and Background: Modern Mercantilism**

After the Second World War, globalization, economic integration at a global scale, accelerated quickly between economies of the world (Grieco, 2000). This move towards "free-trade" as per the definition of Adam Smith, led to global increases in productive efficiency by specialization of nations (Smith, 1776). This can be seen clearly in Figure 1, which shows the continuing decrease in diversity of GDP sectors in the USA over the last 2 centuries. Arguably globalization is one of the key drivers behind technological innovation (Vanham, 2019) as well as a prolonged period of generally peace in the world. Globalization has become the 21st century equivalent of the industrial revolution or the agricultural revolution that came before it.

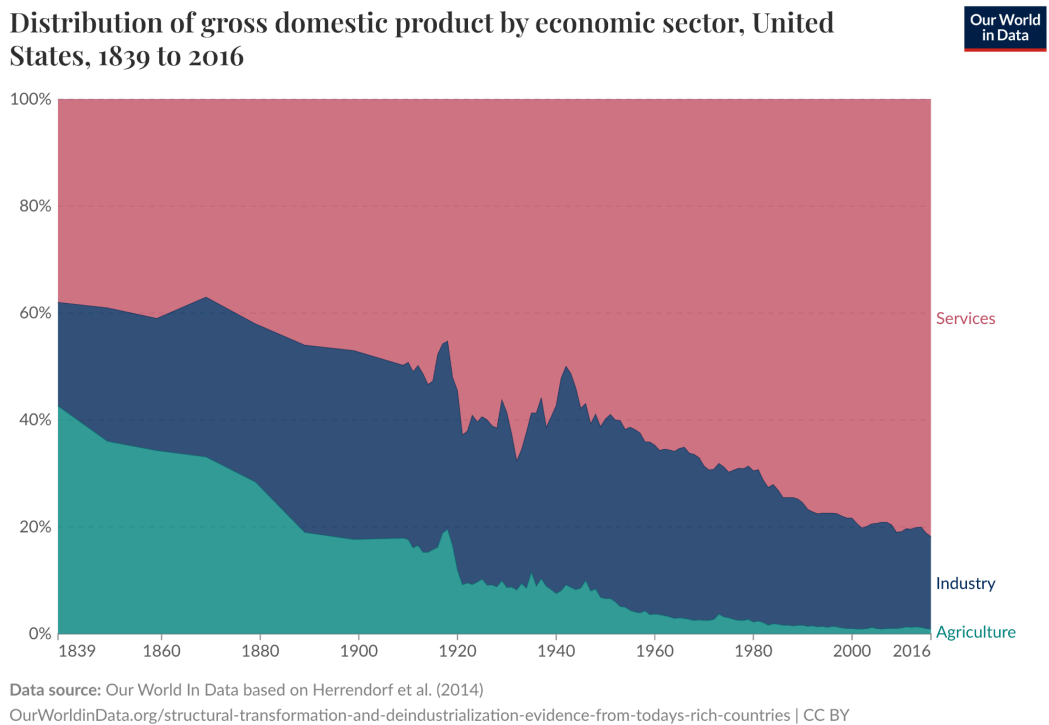

Figure 1 (Our world in data, 2025)

However, over the last couple of years, there has been a noticeable paradigm shift in global economics and trade, from globalization to a much more protectionist approach of nations trying to regain economic autonomy, termed Modern Mercantilism by Bridgewater Associates. Key manifestations include President Trump's Trade War, accelerated institutional friction (WTO), and the China-US Artificial Intelligence (AI) war.

This trend, at a high level, can be attributed to three main events in the modern world: Security Destabilisation (Conflict anxiety and manifestation), AI innovation and Populism (Bradley, 2012). The argument for populism stems from rapid technological advancements at the start of the 2000s *most notably the introduction of smartphones), which led to an orthogonal shift in information consumption funnels. People moved away from relying primarily on large institutions like news networks, shifting instead toward vast ecosystems of short-form user-generated content. This fragmentation democratized information flows but concentrated them into unpredictable, bite-sized formats: the beginning of an attention-economy. It redefined the primary currency governing public engagement. Consequently, political discourse has witnessed a noticeable drift: unconventional, emotionally resonant ideas often gain traction not through rationality or policy expertise alone, but due to their attention-grabbing potential, because attention has become the currency.

In this paper, the predictions (Appendix 1) holistically focus on 4-key areas, with their common denominator being the core dynamics of modern mercantilism :

1. Protectionism driven by Economic Nationalism

This protectionism is explored using the forecast about global subsidies rising for domestic industries over 30% ($P_{YES} = 77\%$) and others, most notably about further tariffs by the U.S. Government ($P_{YES} = 58\%$). From a theoretic framework through a Smithian lens this is highly counterintuitive, as countries are essentially decreasing their working capacity further inside their Production Possibilities Curve (PPC) (Figure 2).

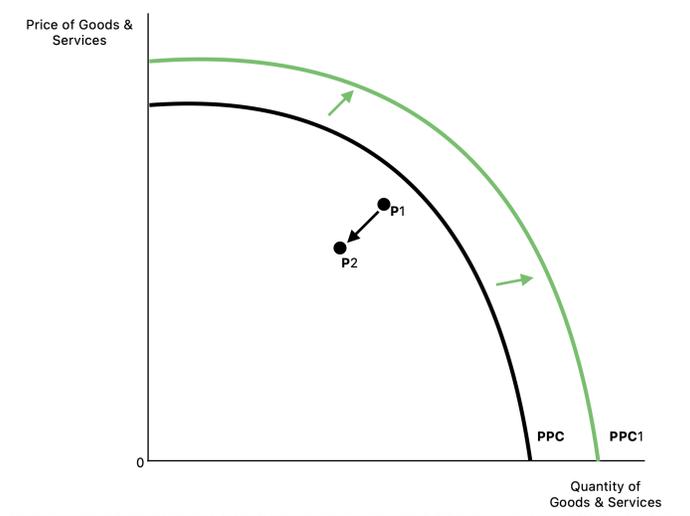

Figure 2

Specialization as a byproduct of globalization would allow the PPC to shift from PPC to PPC1. Arguably, the intent with protectionist policies is that even with the short term lowering in production possibilities, increase in domestic demand will stimulate the domestic market enough to, in the long term, still achieve that push towards PPC1. This is why such policies are likely, as they are explainable and have clear nationalist motivations.

2. Technological Nationalism (gate-keeping)

With the surge in AI technology, most notably the introduction of Large Language Models (LLMs) such as ChatGPT, economies have rekindled an interest in traditional industries and semiconductor appliances.

This is a key economic topic of this decade, and is thus explored using predictions such as 50% of newly installed semiconductor manufacturing capacity (2026–2030) will be located outside East Asia ($P_{\{YES\}} = 68\%$), which suggests protectionism from western governments to aim to be self sufficient in this space.

3. Strategic Competition

Modern Mercantilism is also defined by strategic competition between global powers, where economic tools are used for geopolitical leverage. Examples include outbound investment screening in sensitive technologies like AI/biotech ($P_{\text{YES}} = 68\%$), which reflects the use of economic policy as a strategic filter for national security objectives.

4. Geopolitical fragmentation

Finally, global trade is increasingly shaped by fragmentation into rival blocs, with reduced coordination and increased institutional friction. Predictions such as the decline in G20 intra-bloc trade share by 10% ($P_{\text{YES}} = 65\%$) and the formation of regional blocs that exclude developing nations to counter China ($P_{\text{YES}} = 57\%$) illustrate this movement towards a more fractured, multipolar global economy.

Beyond the standard qualitative economic toolkit, two main elements defined this paper's methodology, providing quantitative analysis of novel data streams. The first one of these is semantic news analysis, which leveraged the recent advancement in the natural language processing space to quantitatively analyze qualitative predictions of experts and reporters. Further, given the domination of the attention-economy, access and subsequently interest in politics and economics has been democratized, thus "crowd" sentiment and beliefs become highly important. Newly formed prediction markets such as Polymarket quantitatively capture this sentiment well, and form a novel methodology of predictive analysis in this paper.

## Analytical Methodology

The Initial Prediction Framework (IPF) aims to determine the Initial Assigned Probabilities (IAPs) of events. Let $E$ denote a binary event, where: $E \in {0, 1}$ The objective of the IPF is to compute the probability that the event occurs, which I define as the Initial Assigned Probability (IAP): $P_0(E = 1) := \text{IAP} \in [0, 1]$ For each event $E$, there exist $n$ Independent Prediction Modules (IPMs). Each module $i$ outputs its own probability estimate: $P_i(E = 1), \quad i = 1, 2, \ldots, n$ Each module is assigned an event-specific weight $w_i$, subject to the following constraints:

$$w_i \geq 0, \quad \sum_{i=1}^{n} w_i = 1$$

These weights represent the importance of each module for this specific events, and are constructed by qualitative human analysis. The IAP is computed as a weighted sum of the individual module outputs, per event:

$$P_0(E = 1) = \sum_{i=1}^{n} w_i \cdot P_i(E = 1)$$

It is also imperative to define that not all of the chosen Events of Interest (EoIs) for forecasting are the same. They are distinguished into two groups:

| Discrete Events | These events have a purely binary outcome, where the event either occurs or does not occur. | The U.S. Government experiences a technical default between 2025 and 2030." | $E \in {0, 1}$ |
|---|---|---|---|
| Continuous Events | These events involve a continuous quantity and resolve based on whether a certain threshold is met or exceeded. | "FDI (Foreign Direct Investment) into China will decrease by more than 20% during 2025–2030 compared to the 2020–2024 average." | Here, the outcome depends on whether the percentage change satisfies: Given, $X$ as the continuous quantity of interest, and $T$ the threshold, it can be formalized as $E = \mathbb{I}[X \geq T] \quad \text{or} \quad E = \mathbb{I}[X \leq T]$ |

Table 2

This splitting in groups is important, as some IPMs, such as semantic news analysis, only perform on discrete and vice versa. In Table 2 each prediction is labelled using this framework, which guides which prediction modules were used.

**Chosen Independent Prediction Modules (IPMs)**

For the purpose of this analysis, a variety of traditional and non-traditional IPMs have been chosen, ranging from semantic analysis to time-series forecasting. Table {{figure number}} provides a brief overview summary of the used IPMs.

| IPM | Description | Reasoning | Used For |
| --- | --- | --- | --- |
| Long-Term Short-Term Memory (LTSM) | Deep learning model for sequence prediction, learning temporal dependencies. | Traditional quant toolkit. One of the best tools for time series forecasting available. | Continuous threshold events. |
| Semantic News Analysis | Semantically analyzing news about events and binarily classifying them. Via k-means, Principal Component Analysis, Zero-Shot Labelling | Probabilistic inference. Semantic analysis of available text encodes latent signals. | Discrete events. |
| Crowdsourced Predictions | Analyzing data on events from crowdsourced betting markets such as Polymarket, Kalshi. Normalized for irrational near-resolution volatility trends. | Betting markets encode public sentiment with efficient aggregation. | Events with available markets, or conditional inference markets where related event outcomes inform probabilities. |
| Macroeconomic Analysis | Heuristic and academic approach to predictions | Traditional forecasting toolkit. Relevant for expressive predictions. | All Events. |

Table 3

## Long-Term Short-Term Memory (LTSM)

As a standardized framework the level of introduction into this module is at a high level. Let $\hat{X}$ denote the LSTM's predicted value at the resolution date, and let $T$ be the binary threshold for the EoI. The absolute distance $|\hat{X} - T|$ is used to derive a confidence weight, such that values far from the threshold imply higher certainty in classification. The probabilistic forecast is then calibrated as: If $\hat{X} \geq T$:

$$P_{\text{Yes}}^{\text{LSTM}} = \sigma(k(\hat{X} - T))$$

If $\hat{X} < T$:

$$P_{\text{Yes}}^{\text{LSTM}} = 1 - \sigma(k(T - \hat{X}))$$

Where $\sigma$ denotes the logistic sigmoid function, and $k$ is a scaling constant to tune the confidence sharpness. $k$ is heuristically defined as $k = 1.5$. This function ensures soft probabilities near the threshold, and higher confidence when the forecast is well above or below. The LSTM module thus informs the Initial Prediction Framework (IPF) by contributing a quantitative, time-series-derived probability for continuous EoIs.

## Semantic News Analysis

Semantic News Analysis (SNA) in this paper refers to computationally understanding sentiment towards the two binary resolutions of an EoI. For ease of modelling, SNA will only consider events of the discrete subgroup exclusively.

At a high level, the SNA module aggregates relevant news articles about EoIs, and first runs them through a Topic Model (BERTopic) to ensure relevancy of articles. The articles are then embedded into a 768-Dimensional vector space using the Sentence Bidirectional Encoder for Transformers (SBERT). These vectors are then analyzed using Principal Component Analysis (PCA) to determine features (dimensions) which most segment the binary outcomes. Further methods include the 2-cluster k-means algorithm and Large Language Model (LLM) zero-shot labelling. Together these form the final probabilities for the SNA IPM.

**News Article Selection**

For news article selection, NewsData.io, NewsAPI and MediaCloud, widely recognized News Application Programming Interface (API) providers were selected to collect relevant news articles from.

For every EoI: a keyword set $\mathcal{K}$ is defined as $\mathcal{K}_i = \text{keyword}_1, \ldots, \text{keyword}_n$, along with a Time window $T$ as $T = [t_0, t_1]$. This is initialized as a query call to each API provider. We collect all articles per EoI within $T$, containing at least 1 keyword from $\mathcal{K}$.

**Relevance Filtering using BERTopic**

To filter the collected articles, the Topical Bidirectional Encoder for Transformers (BERTopic) is used. BERT models are a family of transformer-based language models that use attention mechanisms to capture the contextual meaning of text. This is done by computing a similarity score between an embedding of a generated event summary, $\mathbf{v}_{\text{event}}$, and each news article embedding, $\mathbf{v}_j$. This is done via Cosine Similarity, $S_j$, defined as:

$$S_j = \frac{\mathbf{v}_{\text{event}} \cdot \mathbf{v}_j}{|\mathbf{v}_{\text{event}}| \cdot |\mathbf{v}_j|}$$

The filtered set of news articles, $\mathcal{D}$, is given using a constraint on the cosine similarity threshold, $\tau$:
$$\mathcal{D} = d_j \mid S_j \geq \tau$$

Where $\tau = 0.75$. This threshold was determined heuristically.

**PCA Analysis**

Principal Component Analysis (PCA) is a technique used to explore and reduce the dimensions of high-dimensional vector embeddings, such as those representing news articles. In this paper, PCA is applied to uncover latent features within the embeddings that may influence whether an article leans toward one of the two binary outcomes defined by the discrete EoI.

Initially, a small dataset (±100)[1] of news articles per EoI is manually labelled and attributed towards one of the binary outcomes. The Sentence Bidirectional Encoder for Transformers (SBERT) framework is then applied which gives an 768-dimensional vector embedding: $\mathbf{v}_j \in \mathbb{R}^{768}, \quad j = 1, \ldots, N$

To perform PCA, PC1 has to first be calculated. PC1 represents the direction in the embedding space along which the data varies the most — i.e., it is the axis capturing the greatest possible variance in the dataset. This is done by firstly shifting vectors around the origin (this changes vector location but not their interrelationships):

$$\mathbf{v}'_j = \mathbf{v}_j - \bar{\mathbf{v}}, \quad \bar{\mathbf{v}} = \frac{1}{N} \sum j = 1^N \mathbf{v}_j$$

PCA seeks the direction $\mathbf{u}_1 \in \mathbb{R}^{768}$ *(the first principal component, PC1) that maximizes the projected variance of the data. This is equivalent to solving the following maximization problem:*

$$\mathbf{u}_1 = \arg\max |\mathbf{u}| = 1 \sum j = 1^N \left(\mathbf{u}^\top \mathbf{v}'_j\right)^2$$

---

[1] Note that this is a complexity limitation, as doing a 100 articles will only give n-1 so 99 PC directions, which do not fully exhibit the 768 dimensionality of BERT embeddings.

This finds the unit vector $\mathbf{u}_1$ which captures the greatest variance. By definition, $|\mathbf{u}_1| = 1$; thus, $\mathbf{u}_1$ is an eigenvector of the data's covariance matrix. All subsequent principal components $\mathbf{u}_2, \mathbf{u}_3, \ldots, \mathbf{u}_{768}$ are computed such that they are orthogonal to all previous components: $\mathbf{u}_k^\top \mathbf{u}_j = 0 \quad \text{for all } j < k$

Each $\mathbf{u}_k$ is a unit eigenvector corresponding to eigenvalue $\lambda_k$, and defines PC$k$. The loading score $L_{ik}$ of feature $i$ on PC$k$ is defined as: $L_{ik} = u_{k,i}$

This score indicates the impact of feature $i$ on the construction of PC$k$, and hence its influence on the direction of variance. The variance explained by PC$k$ is given by its eigenvalue $\lambda_k$. Total variance across all PCs is:

$$\text{Var}_{\text{total}} = \sum k = 1^{768} \lambda_k$$

The proportion of variance explained by PC$k$ is:

$$\text{Var}_k = \frac{\lambda_k}{\sum j = 1^{768} \lambda_j}$$

This can be plotted as a scree plot. Dimensionality thus can be reduced to the PCs which retain over 95% variance. The main interest however is to identify the features (dimensions) which best separate the EoI into binary "yes", "no" outcomes as per the manually labelled dataset. To do this, news article embeddings are rotated to PCs as axes: $\mathbf{s}_j = \mathbf{U}^\top \mathbf{v}'_j$

A Fisher Score is then calculated for each principal component (PC), which quantifies the separation between the two classes ("Yes" vs. "No") along that component. Here, $\mu$ denotes the mean and $\sigma$ denotes the standard deviation of the projected values for each class.

$$F_k = \frac{(\mu_{\text{Yes}}^{(k)} - \mu_{\text{No}}^{(k)})^2}{(\sigma_{\text{Yes}}^{(k)})^2 + (\sigma_{\text{No}}^{(k)})^2}$$

The PC$k$ with the highest Fisher Scoring best separates the two classes, denoted as:

$$k^* = \arg\max_k F_k$$

Component $k$ offers maximum separation between the two classes. For this component, loading scores ($L_{ik} = u_{k'i}$) are analyzed. Features (i.e. vector dimensions) with largest absolute loading scores $|L_{ik}|$ are deemed most influential in separating the binary outcomes. This is formalized with a threshold, $\tau_{PCA}$, as:
$$\mathcal{F}_{\text{top}} = \{i \,:\, |L_{ik^*}| \geq \tau_{\text{PCA}}\}$$

The threshold, $\tau_{PCA}$ is selected heuristically as the 95th percentile of loading scores, $\tau_{\text{PCA}} = Q_{95}(|L_{ik^*}|)$. Now, news articles about the EoI which aren't labelled, can also be embedded by SBERT. For a new, unlabeled news article, the SBERT embedding $\mathbf{v}_{\text{new}} \in \mathbb{R}^{768}$ is computed. From this embedding, the values of the selected top features $\mathcal{F}_{\text{top}}$ are extracted:

$$\mathbf{v}_{\text{new}}^{\text{top}} = \{v_{\text{new},i} : i \in \mathcal{F}_{\text{top}}\}$$

Similarly, mean vectors for the Yes and No classes are computed over those top features, from the labeled dataset:

$$\boldsymbol{\mu}\text{Yes}^{\text{top}}, \quad \boldsymbol{\mu}\text{No}^{\text{top}} \in \mathbb{R}^{|\mathcal{F}_{\text{top}}|}$$

Now the distance between the new article and each class mean can be computed:

$$d_{\text{Yes}} = \left|\mathbf{v}\text{new}^{\text{top}} - \boldsymbol{\mu}\text{Yes}^{\text{top}}\right|_2 \quad d_{\text{No}} = \left|\mathbf{v}\text{new}^{\text{top}} - \boldsymbol{\mu}\text{No}^{\text{top}}\right|_2$$

A softmax-like probability can be computed based on inverse distances:

$$P_{\text{Yes}} = \frac{d_{\text{No}}^{-1}}{d_{\text{Yes}}^{-1} + d_{\text{No}}^{-1}} = \frac{d_{\text{No}}}{d_{\text{Yes}} + d_{\text{No}}}$$

And thus: $P_{\text{No}} = 1 - P_{\text{Yes}}$

Once the per-article probabilities $P_{\text{Yes}}^{(j)}$ and $P_{\text{No}}^{(j)}$ are computed for all M unlabeled news articles about a given event, we compute the aggregated event-level probability by taking a weighted average. Let $w_j \in [0, 1]$ denote the weight of article j. The Semantic News Analysis probability for the event, denoted $P_{\text{Yes}}^{\text{event}}$, is computed as:

$$P_{\text{Yes}}^{\text{event}} = \frac{\sum_{j=1}^{M} w_j \cdot P_{\text{Yes}}^{(j)}}{\sum_{j=1}^{M} w_j}$$

Similarly: $P_{\text{No}}^{\text{event}} = 1 - P_{\text{Yes}}^{\text{event}}$

Weighting is done based on the recency of an article. More recent articles receive a higher weighting compared to older ones. This is done by: $w_j = e^{-\lambda t_j} \quad (\text{time decay})$

Where $t_j$ is time in days since posting. This exhaustive framework fully classifies news articles per EoI into two binary groups and estimates the probabilities of each. $\lambda$ is assigned a value of $\dfrac{ln2}{25}$, heuristically.

**K-Means Clustering**

K-Means clustering is an unsupervised learning method by which the news articles per EoI can be clustered into 2 clusters representing binary outcomes of EoIs. It is done initializing arbitrary centroids $c_1, c_2 \in \mathbb{R}^{768}$, assigning vector embeddings to the closest centroid, calculating mean of all assigned embeddings, which becomes the new centroid, and repeating until no more movement. This is another method of clustering to smooth any generalizations from PCA.

First, assign each article, $v_j$ *to a centroid:* $\text{Cluster}(j) = \arg\min k \in {1, 2} |v_j - c_k|_2$

The centroids are then updated by a mean calculation:

$$c_k = \frac{1}{|C_k|} \sum j \in C_k v_j$$

To then workout the assigned probabilities per EoI, we count the number of vectors in each cluster, and weigh by euclidean distance, $d_j$:

$$w_j^{\text{dist}} = \frac{1}{d_j + \epsilon} \quad (\epsilon \text{ is a small constant to avoid divide by zero})$$

and recency score, $t_j$ (age in days): $w_j^{\text{time}} = e^{-\lambda t_j}$

Where the decay constant, $\lambda$ is heuristically defined as:

$$\lambda = \frac{\ln 2}{25} \approx 0.0277$$

meaning an article's weight halves roughly every 25 days. Both weight components have to be normalized using linear normalization:

$$\tilde{w}_j^{\text{dist, time}} = \frac{w_j^{\text{dist, time}}}{\sum k = 1^M w_k^{\text{dist, time}}}$$

These are then equally combined to form:

$$w_j = 0.5 \cdot \tilde{w}_j^{\text{dist}} + 0.5 \cdot \tilde{w}_j^{\text{time}}$$

Let $\mathcal{C}_{\text{Yes}}$ be the set of articles assigned to the "Yes" cluster. Thus the aggregated; assigned weighted probability to the "Yes" outcome is:

$$P_{\text{Yes}}^{\text{event}} = \frac{\sum_{j \in \mathcal{C}_{\text{Yes}}} w_j}{\sum j = 1^M w_j}$$

Given $M$ is the total number of articles. By implication, the complementary probability:

$$P_{\text{No}}^{\text{event}} = 1 - P_{\text{Yes}}^{\text{event}}$$

And hence complete binary probability estimation: $P_{\text{Yes}}^{\text{event}} + P_{\text{No}}^{\text{event}} = 1$ VISUALS FROM CODE

**Zero-Shot Classification**

The zero-shot LLM classification SNA submodule utilizes prompted LLMs to assign news articles into 2 clusters, similar to the k-means algorithm. The used prompt:

---

You are a specialized classification agent.

Your task is to analyze news articles and determine whether each one indicates a "YES" or "NO" outcome for a specific binary event.

Objective: Semantically analyze the news article using your full understanding of language and context. Determine which outcome cluster the article aligns with: YES or NO. Base your decision solely on the article content and its relevance to the event.

Output Format (Strict): Respond with only one of the following, using ALL CAPS with double curly braces: {{YES}} {{NO}}

You must not provide any explanation, commentary, or additional text. Event Context: The binary event is: {{binary_event}}

Classify the Following News Article: {{news_article}}

Enforcement Reminder: Do not explain your choice. Do not output anything except {{YES}} or {{NO}}.

---

This paper used OpenAI's GPT-4o. This is then summed up into a simple ratio of the number of papers in each cluster, which can be formalized as:

$$P_{\text{YES}} = \frac{\sum_{i=1}^{N} \mathbb{1}\left(\text{LLM}(a_i, e) = \text{YES}\right)}{N}$$

and thus $P_{\text{NO}} = 1 - P_{\text{YES}}$ respectively.

The biggest limitation of this approach is the number of articles we can pass through LLMs before it becomes too computationally and financially expensive.

**Final SNA IPM**

Assign heuristic weights $\alpha, \beta, \gamma \geq 0$ with $\alpha + \beta + \gamma = 1$

reflecting qualitative confidence in each submodule. Then the final Semantic News Analysis IPM probability for the event is:

$$\boxed{P_{\text{Yes}}^{\text{SNA}} = \alpha \cdot P_{\text{Yes}}^{\text{PCA}} + \beta \cdot P_{\text{Yes}}^{\text{kmeans}} + \gamma \cdot P_{\text{Yes}}^{\text{ZS}}}$$

Where the weights are determined heuristically.

## Crowdsourcing

The crowdsourcing IPM relies on the "crowd wisdom" anthropological phenomenon which suggests that the aggregate judgment of a diverse group of individuals often surpasses the accuracy of individual experts. Prediction markets like Polymarket are highly efficient aggregators of these dispersed beliefs, due to the inherent financial motivation of the participating "crowd".

For every EoI, if a market exists, we use the probabilities from Polymarket. Let $P_{\text{Poly}}$ denote the implied probability from the market, and $V_{\text{Poly}}$ denote the respective trading volume (via API). The crowd probability is just given by:

$$P_{\text{Crowd}}^{\text{direct}} = P_{\text{Poly}}$$

If the market does not exist, which is very often the case due to the current underground-like culture of prediction markets, heuristically, by factor decomposition available markets are found, $E_1, E_2, \ldots, E_k$ which are closely linked to the EoI. Each of these has a probability, $P_1, P_2, \ldots, P_k$.

Weights $w_i$ are assigned to each proxy market to reflect its qualitative explanatory power toward the EoI, this is done heuristically but also quantitatively by incorporating trading volume. Importantly, since not all sub-events fully span the EoI's scope, the weights are not constrained to sum to 1, but rather:

$$\sum_{i=1}^{k} w_i = \omega \leq 1$$

Here, $\omega$ represents the cumulative explanatory power of the selected sub-events. The residual uncertainty $(1 - \omega)$ captures factors not represented by the selected markets. The inferred crowd probability is calculated as:

$$P_{\text{Crowd}}^{\text{inferred}} = \sum_{i=1}^{k} \frac{w_i}{\omega} P_i$$

This normalization ensures scale-consistency (i.e., the estimate lies in $[0, 1]$), while still allowing for penalization of limited coverage when integrating into the IPF.

To account for unrepresented factors, I apply a penalization to the inferred crowd probability when aggregating into the IPF. Specifically:

$$P_{\text{Crowd}}^{\text{adjusted}} = \omega \cdot P_{\text{Crowd}}^{\text{inferred}} = \sum_{i=1}^{k} w_i P_i$$

For example, if sub-events only capture 70% of the EoI's dynamics, then $\omega = 0.7$, and the inferred crowd probability is scaled accordingly to reflect limited coverage. This penalization ensures that the final aggregation appropriately discounts incomplete market representation during IPF combination.

Lastly, it is necessary to soften each prediction by a time normalization factor, as there is a tendency in crowdsourced markets for there to be irrational volatility by an approaching resolution. This can be expressed as: $P_{\text{Crowd}}^{\text{adjusted}} = P_{\text{Crowd}} \cdot e^{-\lambda t}$

Based on empirical finding, the irrational volatility mostly happens in the last week (7 days) nearing a resolution, so the decay constant is set as $\lambda = \dfrac{\ln 2}{7} \approx 0.099$ thus $P_{\text{Crowd}}^{\text{adjusted}} = P_{\text{Crowd}} \cdot e^{-0.099 \cdot t}$. This constant softens erratic last-minute volatility while retaining most of the crowd signal.

## Macroeconomic Analysis

Macroeconomic analysis was done by application of relevant economic theories to each EoI, and then trying to expand beyond the ceteris paribus convention. An example macroeconomic/trade analysis can be read in the Example Runthrough section.

## Final IPF Calculation

The final IPF for an EoI is an aggregation of all the individual IPMs, heuristically weighted for each EoI individually. This can be denoted as:

$$\boxed{\begin{aligned}P_0(E = 1) = \ &w_{\text{LSTM}} \cdot P_{\text{LSTM}}(E = 1) \\ + \ &w_{\text{SNA}} \cdot P_{\text{SNA}}(E = 1) \\ + \ &w_{\text{Crowd}} \cdot P_{\text{Crowd}}^{\text{adjusted}}(E = 1) \\ + \ &w_{\text{Macro}} \cdot P_{\text{Macro}}(E = 1)\end{aligned}}$$

Subject to: $w_{\text{LSTM}}, w_{\text{SNA}}, w_{\text{Crowd}}, w_{\text{Macro}} \geq 0$ and $w_{\text{LSTM}} + w_{\text{SNA}} + w_{\text{Crowd}} + w_{\text{Macro}} = 1$

## Example Runthrough (Discrete Prediction)

In this section, a practical run through of the IPF will be done on the following prediction, classed as discrete:

> The United States will raise tariffs further on key imports (such as from the EU or Japan) within the next 60 days.

Since it is a discrete event, the algorithmic IPMs that will be run on this EoI include the SNA, crowdsourced information and macroeconomic analysis. LSTM was not run on this EoI,

**SNA**

Articles on the topic were collated using the NewsAPI and NewsData.io with a 60-day backtrack, using relevant keywords. BERTopic run on this dataset collapsing the original set by $\approx 25\%$.

PCA Analysis was now run on a manually labelled dataset of $50$ relevant articles. Figure 3 shows that the $79$ dimensional PC grid had $95$ variance captured by 20 PCs.

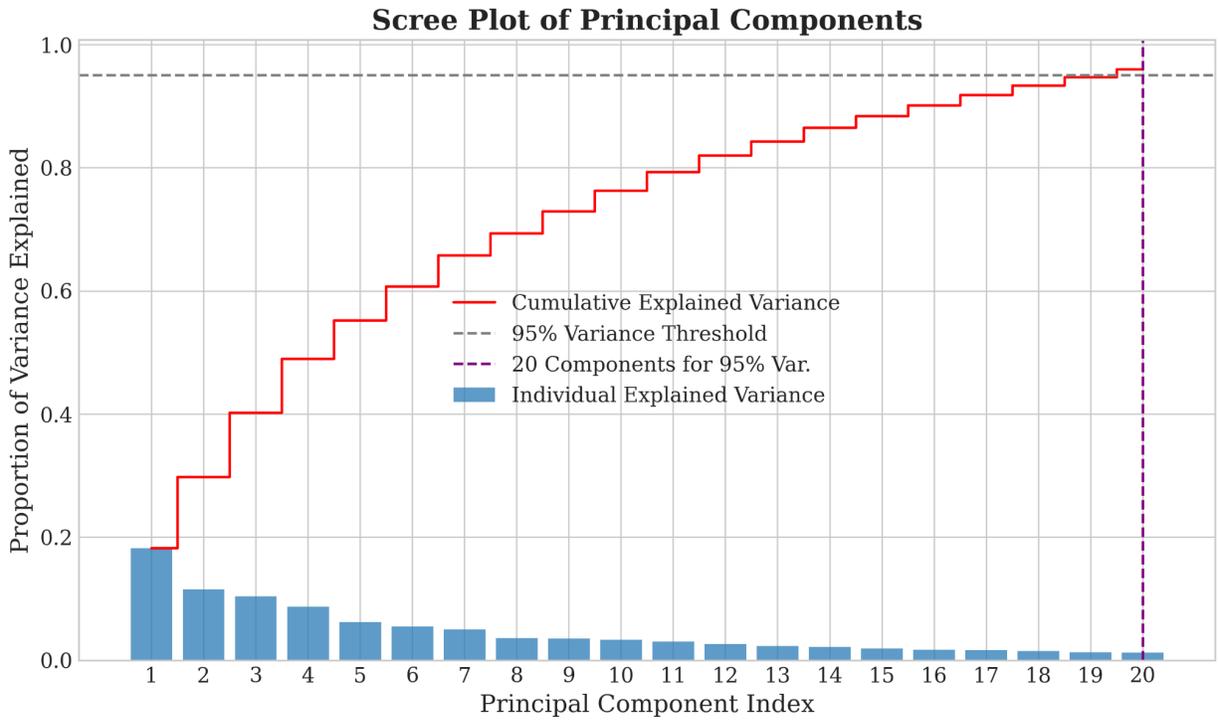

Figure 3

Figure 4 visualizes the top 3 PC dimensions in a dimensionality reduction operation. From the figure it is evident that the "YES" articles cluster together whereas "NO" are outliers.

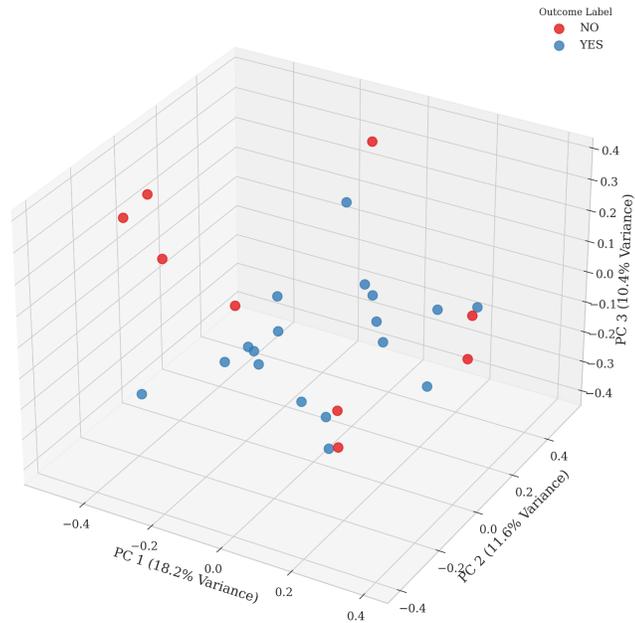

Figure 4

Using the Fisher Score framework, top dimensions were identified which best classify the articles, this can be shown by Figure 5, and the top 3 is made up of PC21, PC3, and PC4 respectively.

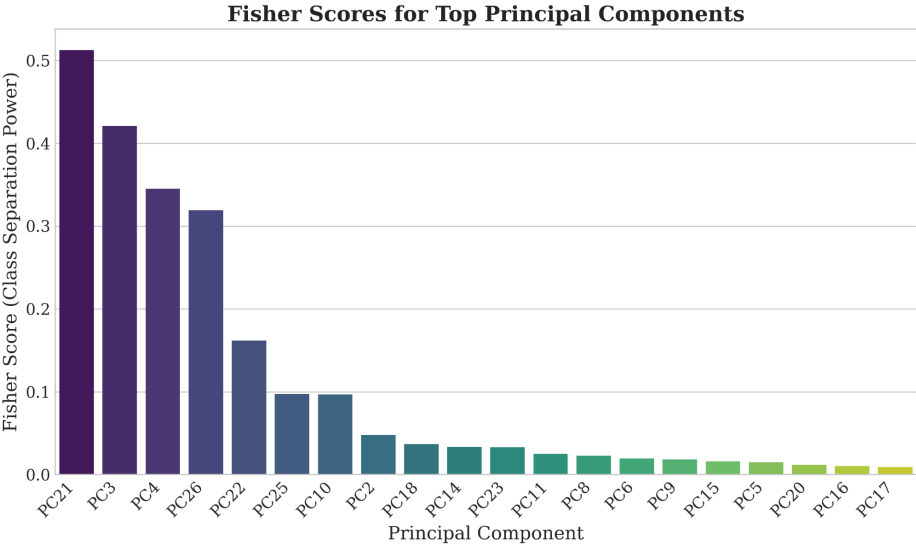

Figure 5

Figure 6 plots the points by dimensionality reduction on the top-3 axes by Fisher Score. From this it is evident that "YES" articles cluster together as do "NO", and prove that these PCs truly separate articles across the binary classification

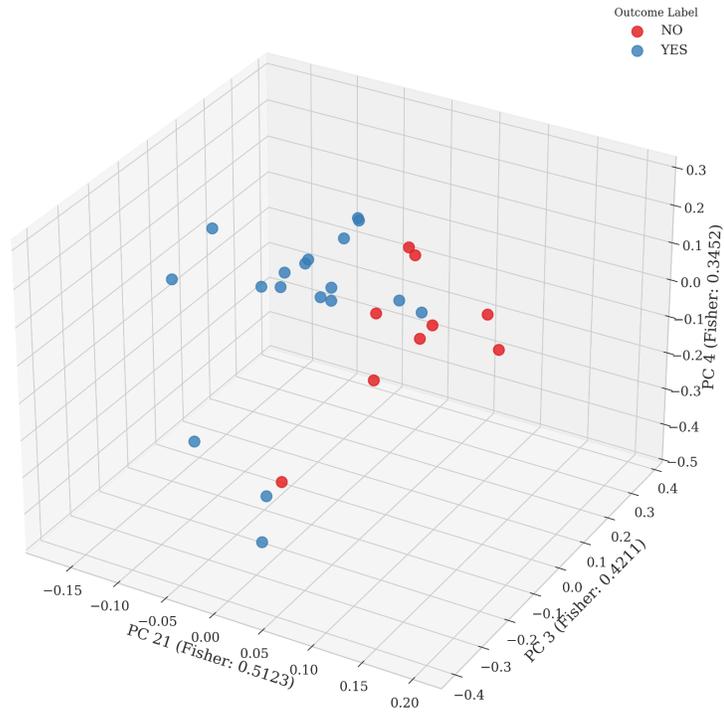

Figure 6

Loading scores (i.e. dimensions) of the SBERT embedding which most contribute to the highest Fisher-rated PCs are by the framework labelled as most important in the binary classification, becoming a mean vector for each class and can now be used to label unseen articles.

Figure 7 shows the different probabilities "YES" stance unlabelled articles take. Note that these are already weighted by the time-decay framework with a half life, $t$, of $t = 25$.

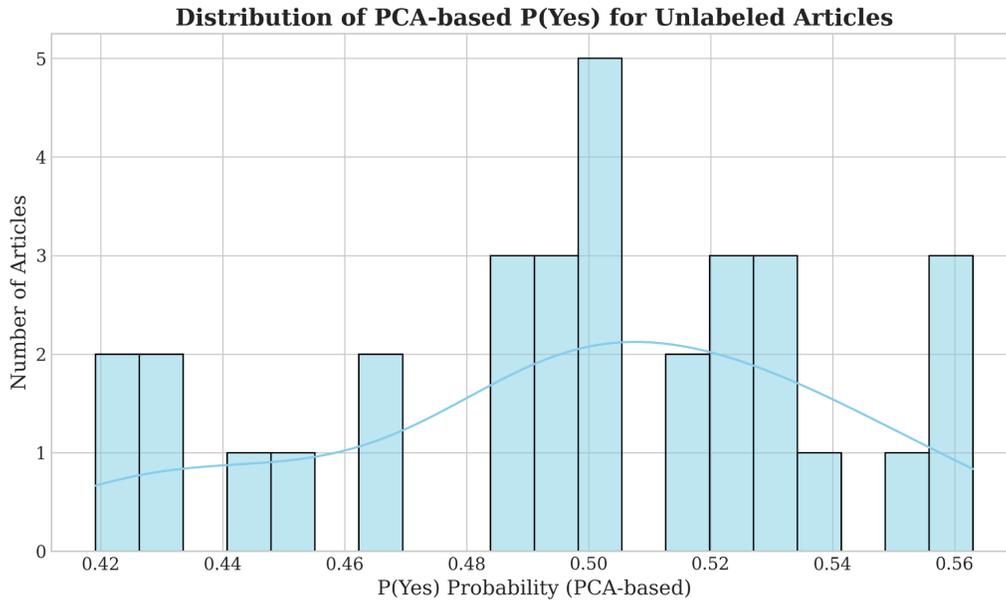

Figure 7

The overall $P_{\text{PCA}}^{\text{YES}}$ of the articles was calculated to $0.4618$.

Figure 8 shows k-means results reduced to two dimensions by PCA[2].

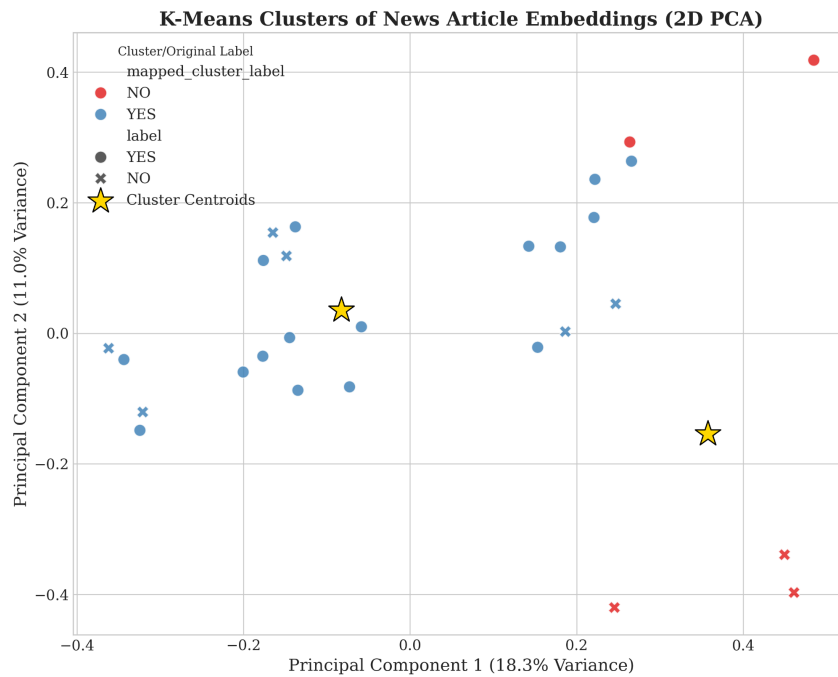

---

[2] Note that PCA here is just a standard framework in this sense for dimensionality reduction for viewing

K-means analysis led to a $P^{\text{YES}}_{(KMEANS)}$ of $0.817$. It is much higher than the PCA probability, but this is most probably a limitation due to a holistic view of the embedding. Thus by heuristic in the final IPM aggregate for SNA for this EoI k-means is going to be given less weight.

Lastly, zero-shot labelling using an LLM gave a $P^{\text{YES}}_{(ZS)}$ of $0.61$.

Heuristically, weightings for each of the 3 components were assigned as follows:

$$\alpha = 0.5, \quad \beta = 0.2, \quad \gamma = 0.3$$

Thus total aggregate $P^{\text{YES}}_{(SNA)}$ is:

$$P^{\text{YES}}_{\text{SNA}} = 0.5773, \ P^{\text{NO}}_{\text{SNA}} = 0.4227$$

**Crowdsourced Information**

3 prediction markets were identified as relevant to the EoI. These are listed below, along with the heuristic weightings, $w$:

| Name | Relevant Probability | TRADING VOL. | Weighting | Resolution Date |
| --- | --- | --- | --- | --- |
| Will the Court Force Trump to Refund Tariffs? | 0.13 PYES | 61230USD | 0.21 | 31.12.2025 |
| Will Trump remove the 10% blanket tariff in 2025? | 0.19 PYES | 29074USD | 0.2 | 31.12.2025 |
| Will Trump lower tariffs on Mexico by December 31? | 0.83 PYES | 6965USD | 0.05 | 31.12.2025 |

Thus $P^{\text{YES}}_{\text{[\texttt{yes?}]}crowd}$ is given by:

P^inferred_Crowd = 0.2322
**P^adjusted_Crowd = 0.1068**

Note that this is low because of the lack of full explanatory power. Its impact will be fully downweighted in the final IPF aggregate due to this.

**Macroeconomic Analysis**

From a macroeconomic standpoint there are many factors that come to play when determining the likelihood of further tariffs presented by the U.S.

From a Smithian perspective, free trade (absence of government intervention) theoretically maximizes production efficiency, increases competition and lowers prices for consumers, as can be seen in Figure 9a by the benefit for consumers purchasing at $P_w$ compared to $P_{\text{dom}}$, leading to imports $Q_4 - Q_1$. If tariffs are introduced, $P_w$ shifts to $P_{w+t}$, which decreases the quantity of imports. This is not beneficial to the consumer who will now have to pay the higher price for the good. In the context of the EoI, we are looking at a potential further raise of tariffs for key economic partners. This would lead to the new price $P_{w+t+T}$ as seen on Figure 9b. The marginal gain on government revenue is low, whereas the marginal welfare loss is large. Such a situation would further impact consumer prices, and could become inflationary. Thus from a short-term economic perspective it is not very feasible, and counterintuitive.

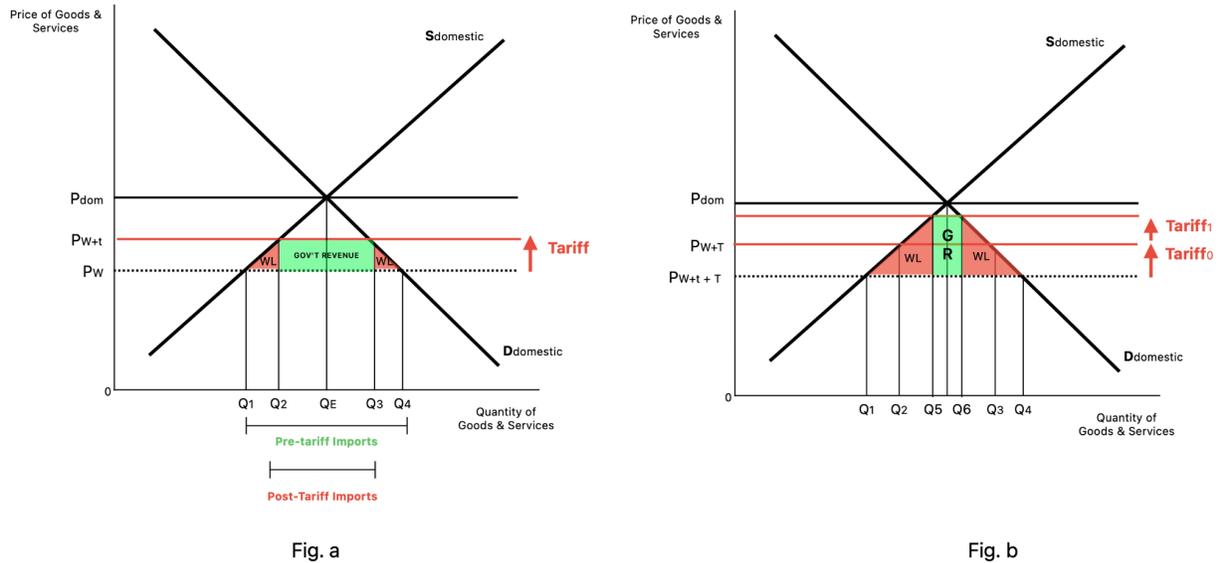

Figure 9

However, it is important to take this potential trade policy in context of President Trump's second term, where his policies are presented as long term, aiming to revitalize U.S. industries by increasing domestic supply, which in turn can lead to domestic economic growth. Paradoxically, it is also important to note Trump's push on lowering interest rates by the Fed, which would be less likely to occur given further inflationary cause-and-effect given these policies.

Even if by standard theoretic economic trade analysis further interest rates are not very plausible, given the broad nature of the EoI, it may still be very likely that there will be increases in tariffs to key economic partners, as beyond the ceteris paribus convention these can be explained as "cautionary" tariffs to strengthen Trump's position in negotiations with other countries, and receive better terms, as seen by the recent US-EU trade deal.

Thus by macroeconomic analysis, $P^{\text{YES}}_{\text{macro}}$ is given a 56% probability.

**Total Prediction from Discrete**

Thus the total prediction is:

$$P^{\text{YES}} = 0.5 \cdot P^{\text{YES}}_{\text{SNA}} + 0.1 \cdot P^{\text{YES}}_{\text{crowd}} + 0.4 \cdot P^{\text{YES}}_{\text{macro}}$$

$$= 0.5 \cdot 0.5773 + 0.1 \cdot 0.1068 + 0.4 \cdot 0.56$$

$$P^{\text{YES}} = 0.52333$$

$$P^{\text{NO}} = 1 - P^{\text{YES}} = 1 - 0.52333 = 0.47667$$

$$P^{\text{YES}} = 0.52333, P^{\text{NO}} = 0.47667$$

**Reference List**

Bradley, B. (2012). *Post-war European Integration: How We Got Here*. [online] E-International Relations. Available at:
https://www.e-ir.info/2012/02/15/post-war-european-integration-how-we-got-here/.

Grieco, J. (2000). *The International Political Economy Since World War II Prepared for the CIAO Curriculum Case Study Project*. [online] Available at:
https://www.files.ethz.ch/isn/6843/doc_6845_290_en.pdf.

Smith, A. (1776). *The Wealth of Nations*. London: W. Strahan and T. Cadell.

Vanham, P. (2019). *A Brief History of Globalization*. [online] World Economic Forum. Available at:
https://www.weforum.org/stories/2019/01/how-globalization-4-0-fits-into-the-history-of-globalization/.

World In Data (n.d.). *Shares of GDP by economic sector*. [online] Our World in Data. Available at: https://ourworldindata.org/grapher/shares-of-gdp-by-economic-sector.

**Appendix 1: 20 Binary Forecasts**

| Forecast | Resolution Criteria | Event Type | P YES | P NO |
|---|---|---|---|---|
| 50% of newly installed semiconductor manufacturing capacity (2026–2030) will be located outside East Asia. | SEMI or Gartner fab capacity reports (wafer capacity or new fab count). | Continuous | 68% | 32% |
| FDI into China will decrease >20% (2025–2030) vs 2020–2024 average. | UNCTAD or China Ministry of Commerce FDI data. | Continuous | 74% | 26% |
| Global subsidies for domestic industries will rise ≥30% (2026–2031 vs 2020–2025). | OECD or IMF subsidy totals. | Continuous | 77% | 23% |
| U.S. government experiences technical default between 2025–2030. | Treasury or rating agency reports (missed payments or downgrade due to default). | Binary | 19% | 81% |
| USA leave the WTO | Official White House Statement | Binary | 12% | 88% |
| India joins a major regional trade bloc | Official bloc membership + | Binary | 63% | 37% |
| (excluding SAARC) that excludes China by 2030. | explicit exclusion of China. | | | |
| Two G7 nations implement "friend-shoring" policies targeting supply chains from China by 2029. | Official incentive or subsidy policies documented. | Binary | 75% | 25% |
| US/EU/UK require "security reviews" for outbound cloud/AI services to China before 2028. | Binding regulations or policies mandating such reviews. | Binary | 66% | 34% |
| Global FDI into China declines ≥25% (vs 2022) in any year by 2028. | UNCTAD annual FDI data. | Continuous | 71% | 29% |
| EU denies ≥€2B in foreign high-tech/energy investments on security grounds by 2029. | EU Commission or national decisions to block investments. | Continuous | 54% | 46% |
| The United States will raise tariffs further on key imports (such as from the EU or Japan) within the next 120 days. | US Government | Binary | 52% | 48% |

| Question | Resolution Source | Type | Yes | No |
|---|---|---|---|---|
| EU Parliament adopts "Buy European" procurement policy for strategic goods by 2028. | Legislation or resolution specifying domestic sourcing requirements. | Binary | 61% | 39% |
| EU export restrictions on "info security/cryptanalysis" items rise ≥30% by 2028 (vs 2023). | Updates to EU dual-use export control lists. | Continuous | 58% | 42% |
| U.S. revokes China's MFN trade status by July 2027. | Congressional or Executive action formally revoking MFN. | Binary | 33% | 67% |
| WTO trade disputes filed increase ≥20% in any year by 2029 vs 2022. | WTO annual dispute case filings. | Continuous | 67% | 33% |
| Global trade volume growth <2% per year, 2026–2028. | WTO trade volume data, annualized growth rates. | Continuous | 49% | 51% |
| China signs zero-tariff deal with 30+ African middle-income countries by 2027. | Official trade agreement + tariff schedule implementation. | Continuous | 59% | 41% |
| Africa's share of global exports remains <3.5% | WTO or IMF export share data. | Continuous | 72% | 28% |
| through 2030 despite AfCFTA. | | | | |
| Two G20 nations implement outbound investment screening for AI/biotech by 2029. | National laws or official policies enacted. | Continuous | 68% | 32% |
| G20 intra-bloc trade share declines ≥10% by 2029 (global trade fragmentation). | WTO or IMF data on G20 trade shares. | Continuous | 65% | 35% |
| A major regional trade bloc will exclude at least one developing nation by 2030 to challenge Chinese economic influence. | Trade Bloc Announcement | Binary | 57% | 43% |

Table 1